# Hα spectropolarimetry of the B[e] supergiant GG Carinae*

A. Pereyra[1], F. X. de Araújo[1], A. M. Magalhães[2], M. Borges Fernandes[3], and A. Domiciano de Souza[3]

[1] Observatório Nacional, Rua General José Cristino 77, São Cristovão, 20921-400, Rio de Janeiro, Brazil
[2] Departamento de Astronomia, IAG, Universidade de São Paulo, Rua do Matão 1226, São Paulo, SP, 05508-900, Brazil
[3] UMR 6525 H. Fizeau, Univ. Nice Sophia Antipolis, CNRS, Observatoire de la Côte d'Azur, France

**ABSTRACT**

*Aims.* We study the geometry of the circumstellar environment of the B[e] supergiant star GG Car.
*Methods.* We present observations acquired using the IAGPOL imaging polarimeter in combination with the Eucalyptus-IFU spectrograph to obtain spectropolarimetric measurements of GG Car across Hα at two epochs. Polarization effects along the emission line are analysed using the $Q$–$U$ diagram. In particular, the polarization position angle (PA) obtained using the line effect is able to constrain the symmetry axis of the disk/envelope.
*Results.* By analysing the fluxes, GG Car shows an increase in its double-peaked Hα line emission relative to the continuum within the interval of our measurements (∼43 days). The depolarization line effect around Hα is evident in the $Q$–$U$ diagram for both epochs, confirming that light from the system is intrinsically polarized. A rotation of the PA along Hα is also observed, indicating a counter-clockwise rotating disk. The intrinsic PA calculated using the line effect (∼85°) is consistent between our two epochs, suggesting a clearly defined symmetry axis of the disk.

## 1. Introduction

Early studies of GG Car (Pickering 1896a, 1896b) showed that it displayed a peculiar spectrum with strong emission lines. McGregor et al. (1988) identified CO absorption bands in GG Car along with characteristics of the B[e] phenomenon, which include strong Balmer lines in emission, low excitation permitted emission lines (e.g., FeII), forbidden emission lines (e.g., [FeII] and [OI]), and strong infrared excess (Zickgraf 1998). Lamers et al. (1998) classified this object as a Galactic B[e] supergiant (sgB[e]) based on estimates of its effective temperature and luminosity.

Photometric evidence of GG Car as an eclipsing binary was found by Gosset et al. (1984), with a period of 31.02 or 62.04 days and a typical amplitude Δmag∼0.5. The first period was confirmed from radial velocities of several emission and absorption lines (Hernández et al. 1981; Gosset et al. 1985). The presence of HeI in emission suggests that one of the components is a B0-B2 star (Lopes et al. 1992).

The distance of GG Car from interstellar NaI lines seems consistent with ∼2.4kpc (Lopes et al. 1992) but may have been overestimated by a factor of 2 (Lopes 2009, private comm.). The effective temperature of GG Car was estimated to be $T_{eff}$∼1.6×10$^4$ K and its luminosity $log(L/L_\odot)$=5.18 (McGregor et al. 1988). Mass estimates for GG Car include $M_{ZAMS}$=25M$_\odot$ and $M_{B[e]}$=23M$_\odot$ (de Araújo & de Freitas Pacheco 1994; de Freitas Pacheco 1998).

Gnedin et al. (1992) detected optical polarization variability (ΔP∼0.4−0.9%) in GG Car. Optical broadband polarization was previously reported by Klare & Neckel (1977) and Barbier & Swings (1982). The long-term polarization variability indicates that GG Car does have intrinsic polarization, which probably originated in light scattering off one of the binary components or a variable circumstellar disk/wind.

Concerning spectral features, there is strong variability in GG Car, especially regarding whether or not P-Cygni profiles are detected (Smith 1955; Swings 1974; Lopes et al. 1992), which is indicate of again significant wind variability. In addition, Machado et al. (2004) using high resolution spectra confirmed the binarity and intrinsic variability of GG Car. Groh et al. (2007) detected Paγ and HeI 10830 Å in emission along with several features produced by FeII, MgII, and CI in its near-infrared spectrum. In addition, GG Car appears to display $^{13}$CO band emission in its mid-infrared spectrum, indicating that it is indeed a sgB[e] star (Kraus 2009) based on the enrichment of its circumstellar matter by $^{13}$C.

Spectropolarimetry provides insight into stellar envelopes where scattering opacities exist without the need to resolve the envelope (Magalhães 1992; Magalhães et al. 2006). It provides information about the envelope geometry and the structure of the line formation region. In this work, we report the first spectropolarimetric observations across Hα of GG Car. The observations and data reduction are presented in Sect. 2. The results are shown in Sect. 3 and the conclusions are drawn in Sect. 4.

## 2. Observations

The observations were performed during two runs in 2006 April and May using the 1.6m telescope at the Observatório

---
*Send offprint requests to*: A. Pereyra, e-mail: `pereyra@on.br`
* Based on observations obtained at the *Observatório do Pico dos Dias*, LNA/MCT, Itajubá, Brazil.

do Pico dos Dias (OPD-LNA), Brazil. A log of the observations is shown in Table 1. We used IAGPOL, the IAG imaging polarimeter (Magalhães et al. 1996, Pereyra 2000), installed in the Eucalyptus-IFU spectrograph (EIFU, de Oliveira et al. 2003). EIFU is an integral field unit consisting of an array of 32 × 16 fibers of 50 µm, which covers a field of 30″ × 15″ on the sky with a scale of 0.93″ per pixel. The detector used was a Marconi 2048 × 4608 pixel back-illuminated CCD with 13.5 µm$^2$ per pixel. EIFU uses a 600 l/mm grating. It provides a spectral range of ∼600 Å around Hα and a resolution $R$=4000, or ∼0.3 Å per pixel.

We used IAGPOL in linear polarization mode. It employs a calcite Savart plate as an analyzer and an achromatic $\lambda/2$-waveplate as a retarder above the Savart plate. Each measurement consisted of eight waveplate positions separated by 22°.5. The IAGPOL+EIFU setup (Pereyra et al. 2009) divides the light collected by the telescope into two orthogonal polarization beams, which are projected onto the EIFU fiber array. Spectropolarimetry can then be performed using the ordinary and extraordinary stellar beams, which yield -o and -e spectra at each waveplate position. For an example of the reconstructed polarized images, seen by the fiber array, we refer to Fig. 1 of Pereyra et al. (2009).

We used standard *IRAF*[1] procedures to complete IFU reduction at each waveplate position image including bias and flatfielding corrections along with wavelength calibration. A specially developed routine was then used to extract and stack fibers for the -o and -e beams and the sky region. The optimum aperture radius for the fiber extraction was selected by minimizing the polarimetric errors. Therefore, three spectra were constructed at each waveplate position: O($\lambda$), E($\lambda$), and sky($\lambda$). Special care was taken to remove cosmic rays.

For both epochs, the -o and -e spectra were constructed using an aperture radius of 4″ on the fiber plane and by stacking 49 fibers around the -o and -e centroids from the 2-dimensional IFU images. The sky spectrum was constructed by stacking fibers away from the -o and -e beams. The sky was then subtracted from the -o and -e spectra at each waveplate position image. Several sky positions within the reconstructed image were tested with similar results. Typical signal-to-noise ratios provided by our setup are higher than 100 (Table 1).

After that, we used the SPECPOL[2] package to construct the flux, polarization, and polarization position angle (PA) spectra for a proper binning. To optimize the signal-to-noise ratio, this package allows binning of the spectra using a variable bin size set by a selected and constant polarization error per bin. The errors are obtained from the residuals of the observations at each waveplate position image with respect to the expected cosine curve. In general, they are consistent within the uncertainties determined by photon statistics. The calibration process, which includes observations of polarized and unpolarized standard stars, was similar to previous runs performed with this setup (Pereyra et al. 2009).

---

[1] *IRAF* is distributed by the National Optical Astronomy Observatory, which is operated by the Association of Universities for Research in Astronomy, Inc., under cooperative agreement with the National Science Foundation.

[2] The original version of SPECPOL was written by A. Carciofi and is available within PCCDPACK (Pereyra 2000).

**Table 1.** Log of observations and results for GG Car.

| Date | IT$^a$ (s) | S/N$^b$ | $P_{\rm obs}$$^c$ (%) | PA$_{\rm obs}$$^c$ (°) | PA$_{\rm int}$ (°) |
|---|---|---|---|---|---|
| 2006 Apr. 05 | 4800 | 104 | 1.6 (0.1) | 85 (2) | 84 (6) |
| 2006 May 18 | 4800 | 154 | 1.5 (0.1) | 84 (2) | 85 (4) |

Errors in parenthesis. $^a$ Total integration time (8 × individual integration time by waveplate position); $^b$ signal-to-noise ratio per pixel for the unbinned spectra; $^c$ mean values for the full spectrum (6300–6800 Å) binned with a polarization error (per bin) of 0.1%. The quoted polarization error is the average error per bin and the PA error is given by 28.65×$\sigma_P/P$ (Serkowski 1974).

**Table 2.** Temporal evolution in the multicolor polarization of GG Car from the literature.

| Date/epoch | Band | $P$ (%) | PA (°) | ref.$^a$ |
|---|---|---|---|---|
| 1973–1975 | V | 2.2 (0.1) | 99.3 | 1 |
| 1980 Mar. | U | 0.8 (0.1) | 99.0 | 2 |
|  | B | 1.6 (0.1) | 99.0 | 2 |
|  | V | 1.65 (0.4) | 95.0 | 2 |
| 1989 May | U | 1.6 (0.1) | 100 | 3 |
|  | B | 2.5 (0.1) | 100 | 3 |
|  | V | 2.1 (0.1) | 100 | 3 |
|  | R | 1.9 (0.1) | 100 | 3 |

Errors in parenthesis. $^a$ (1) Klare & Neckel (1977), mean of two measurements. It is not clear which broadband filter was used. We assumed optical V broadband for simplicity; (2) Barbier & Swings (1982), mean of five measurements and after taking into account an interstellar contribution; (3) Gnedin et al. (1992). The PA quoted is the mean value including all filters as reference.

## 3. Results

Table 1 shows the observed[3] continuum polarization ($P_{\rm obs}$ and PA$_{\rm obs}$) for GG Car integrated along the full range of our spectra (6300–6800 Å) between 2006 Apr. and May. Within the errors, the continuum polarization level and its PA were unchanged over ∼43 days. Nevertheless, by comparing with historical optical broadband polarization data (see Table 2), significant variability is found.

Our Hα spectropolarimetry is indicated in Fig. 1 for our two epochs. The spectral range of the top spectra is only 45Å for a clearer visualization of the line feature. The spectra are binned using a variable bin size with a constant polarization error (per bin) of 0.1%. Significant changes in the polarization level and its PA, indicative of a detected line effect, are observed across the Hα emission for both dates, which immediately suggests that intrinsic polarization in GG Car is present.

The lower plots of Fig. 1 are the $Q-U$ diagrams showing the line effect that is evident around the emission line between 6555 and 6568 Å. This wavelength range is also indicated (by vertical lines) in each spectrum (Fig. 1, top) for a clearer comparison. In 2006 Apr., the line effect (black dots in the $Q-U$ diagram) is consistent with depolarization showing a linear excursion that points approximately to the $(Q,U)$ coordinates origin. On the other hand, the 2006 May data exhibit also depolarization but with a clearly defined PA rotation, which resembles a mix between a loop and a linear excursion. The non-detection of the loop in the 2006 Apr. data may be the result of the lower signal-to-noise ratio in this measurement (Table 1). In general, this depolarization level can be used to estimate the foreground interstellar polarization (ISP; e.g., Schulte-Ladbeck & Clayton 1993) along the line of sight of GG Car. This assumption is valid if the line emission itself is unpolarized close to the line center. Consistent with this, in both epochs, the highest depolarization occurs exactly at the line center (∼6561 Å) and at the same level within the errors, the average being $(Q,U)_{\rm ISP} = (-0.67\%\pm0.10\%, +0.14\%\pm0.10\%)$ or $P_{\rm ISP}$∼0.69% at PA=84°.

---

[3] In general, the observed polarization is a combination of intrinsic and interstellar polarization.

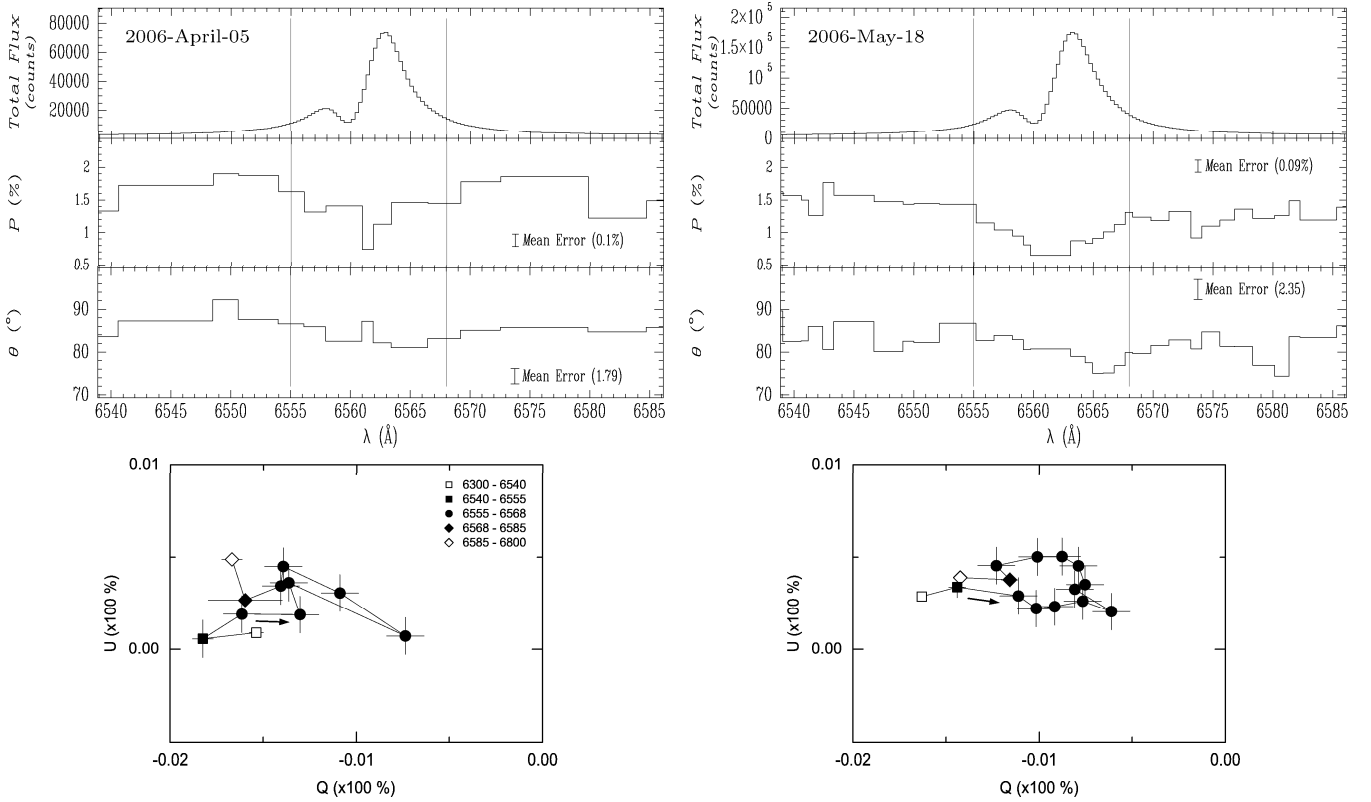

**Fig. 1.** Spectropolarimetry of GG Car around Hα on 2006 Apr. 05 (left) and 2006 May 18 (right). *Top*: The vertical lines indicate the region (6555–6568 Å) where the line effect is evident. The spectra show the total flux (top), polarization (middle), and polarization PA (bottom) binned using a variable bin size with a constant polarization error (per bin) of 0.1%. *Bottom*: $Q-U$ diagram showing the line effect (black dots). The direction in which the wavelength increases is also indicated (arrow) along with the mean values for the blue (white square, 6300–6540 Å) and red (white diamond, 6585–6800 Å) continuum. The mean values for the regions immediately before (black square, 6540–6555 Å) and after (black diamond, 6568–6585 Å) the line effect are also shown.

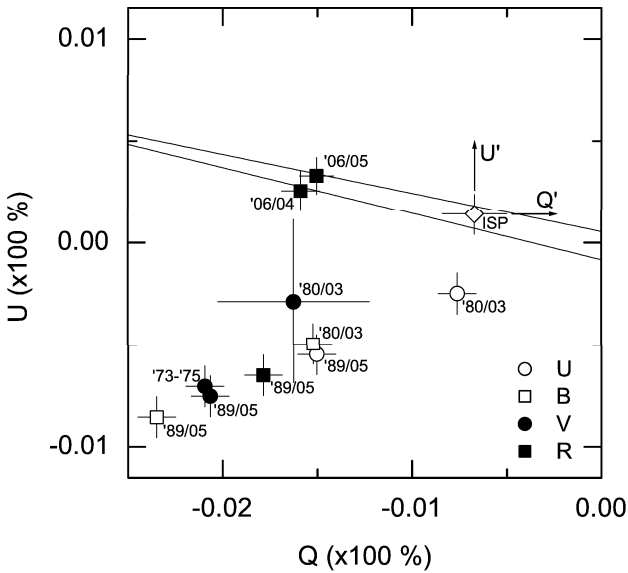

**Fig. 2.** $Q-U$ diagram for GG Car showing the intrinsic polarization PA (solid lines) from the Hα line effect in 2006 Apr. and 2006 May. The continuum polarization around Hα (same symbol as R filter) is also shown along with the historical temporal multicolor polarization from Table 2. The mean interstellar polarization obtained from the depolarization at the line center is also indicated (white diamond). The arrows indicate the reference system ($Q'$, $U'$) corrected for the ISP.

Interestingly, the direction in which the wavelength increases around the line effect (indicated by the arrows in the $Q-U$ diagrams) has the same sense in our two epochs. This direction is such that if a disk exists around GG Car it must rotate in a counter-clockwise direction as seen by an observer on Earth (Poeckert & Marlborough 1977). In addition, a small rotation (of a few degrees) in the PA for the integrated continuum polarization seem to be present with the bluer continuum being slightly more polarized. This effect is more consistent in the 2006 May data, which has a higher signal-to-noise ratio.

To determine the intrinsic PA (PA$_{\rm int}$), we completed a linear fit to the points that display the line effect in the $Q-U$ diagram. Formally, the slope of the fit yields two possible solutions (PA$_{\rm int}$ and PA$_{\rm int}$+90°). In general, this ambiguity is solved by choosing the solution to be more consistent with the relative positions of the ISP and the observed polarization in the $Q-U$ diagram (Schulte-Ladbeck et al. 1994).

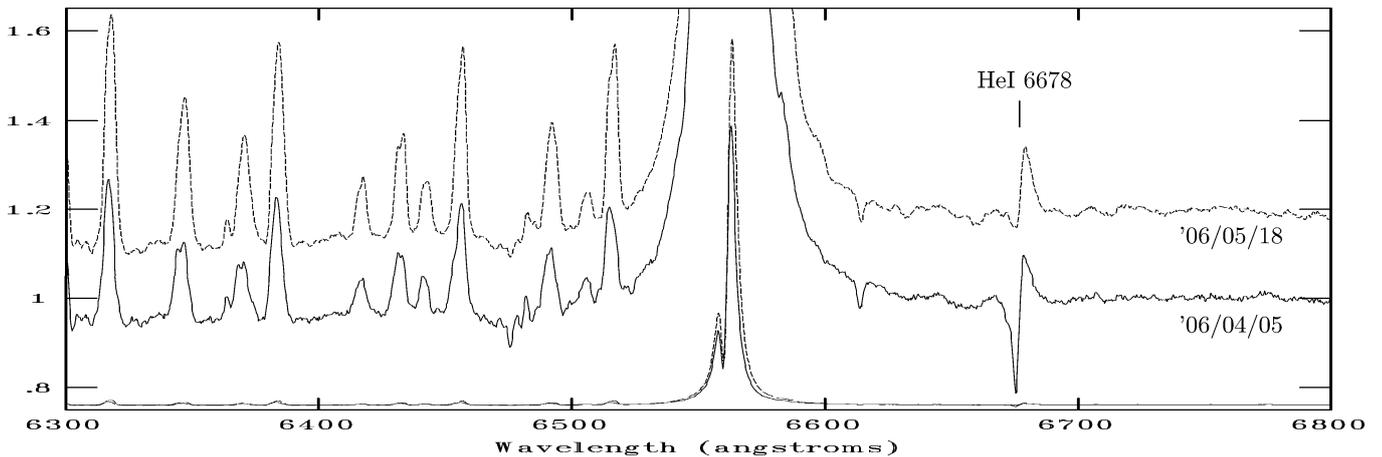

**Fig. 3.** Comparison of the full EIFU spectra around Hα from GG Car between 2006 Apr. 05 (solid line) and 2006 May 18 (dashed line). The fluxes were normalized to the continuum in each case.

Figure 2 shows the observed continuum polarization and the $PA_{int}$ directions obtained from the line effect for our two epochs (Table 1). The ISP toward GG Car obtained by us using the depolarization in the line center is also shown. Analysing this diagram it seems clear that the solutions with $Q' < 0$ for $PA_{int}$ are more likely. These solutions are given in the last column of Table 1. As we can see, the $PA_{int}$ values are practically identical (∼85°) during our two epochs considering the individual errors.

We can also obtain information about the intrinsic polarization using the observed continuum polarization ($P_{obs}$ in Table 1) corrected for the ISP that had already been computed previously. In this case, the mean intrinsic polarization averaged between our two epochs is 0.9% at PA=84°. Interestingly, the last value is exactly the same $PA_{int}$ computed using the line effect with the intrinsic polarization being parallel to the ISP. The $PA_{int}$ computed by the above two methods represents a clearly defined symmetry axis for the disk. If this is true, the $PA_{int}$ will be perpendicular or parallel to the disk plane depending on whether we have an optically thin or optically thick disk, respectively (Vink et al. 2005). Consistent with this issue, VLTI/MIDI interferometry (Domiciano de Souza 2009, private comm.) also detects a disk plane close to being perpendicular to our $PA_{int}$ that is indicative of an optically thin disk.

As we noted previously, the historical broadband polarization in GG Car exhibits significant variability. To facilitate the comparison with our data, we also plotted these data in the $Q-U$ diagram (Fig. 2). Our data seem to indicate that the intrinsic polarization measured by the line effect or by the ISP corrected continuum is of the same origin, which we believe to be associated with the disk. Therefore, assuming that the geometry of the disk is constant in time, the $Q-U$ diagram indicates that the historical continuum variability must also have an additional component. In this context, the observed temporal variation between 1973-1989 and our data can be attributed to time-dependent physical properties of the wind of GG Car. Possible mechanisms that can explain this variation include a rise in the stellar mass-loss rate as in the sgB[e] star HD 34664 (Schulte-Ladbeck & Clayton 1993), or eventual aspherical blowouts of material as in the peculiar B[e] star HD 45677 (Patel et al. 2006).

Additional information about the GG Car binary system can be found by analysing the flux spectra. Figure 3 shows the comparison of the full EIFU spectra normalized to the continuum at both epochs. As noted previously (Fig. 1), in 2006 April Hα appeared to have a double-peaked profile, the central dip being above the continuum level and the redder peak being more intense (red-to-blue peak ratio, $R/B$ ∼3.6 and red-peak-to-continuum ratio, $R/C$ ∼30). On the other hand, the 2006 May data exhibit also a similar double-peaked emission ($R/B$ ∼3.8) but with a significant increment relative to the continuum, $R/C$ ∼39. Interestingly, if we consider the periodic light curve of Gosset et al. (1985) for the binary system, our epochs corresponded to maximum ($\phi$=0.54) and minimum ($\phi$=0.92) brightness, respectively. Therefore, if the continuum flux variability is caused mainly by the eclipse of the companion and Hα is formed in a rotating disk around the primary, the emission must exhibit a corresponding enhancement in its minimum brightness but maintain the relative intensities of its Hα emission components, as we have observed. However, HeI 6678 Å displayed a clearly defined P-Cygni profile in 2006 April, its absorption component then practically disappearing into the continuum level in 2006 May. This variation is inconsistent with an occultation by the companion at minimum brightness and suggests a system with a circumbinary disk. More measurements are needed to clarify this issue.

## 4. Conclusions

For the the first time, we have presented spectropolarimetry data of the B[e] supergiant GG Car. The continuum polarization around Hα was constant during the interval of our measurements (∼43 days). Nevertheless, by analysing the $Q-U$ diagram, polarization line effects on the line profile were observed with a recurrent depolarization along with a polarization loop in the last epoch associated with a counter-clockwise rotating disk. The depolarization effect was used to compute the foreground interstellar polarization at the line center. Interestingly, the position angle of the intrinsic polarization computed using either the line effect or the continuum corrected for the foreground polarization are practically the same (PA ∼85°). This symmetry axis can be used to constrain the position angle of the disk projected on the sky. Correlations with the disk plane detected by interferometry suggest that an optically thin disk is present in GG Car.

An enhancement in the double-peaked Hα emission profile was also observed along with a change in the absorption component of the P-Cygni profile in HeI 6678 Å. The Hα variation seems to be correlated with the maximum and minimum phases of the eclipsing system but does not provide conclusive evidence of a disk around the primary component or a circumbinary disk. This work has demonstrated the potential of using spectropolarimetry to determine the geometry of the wind/disk environment associated with B[e] supergiant stars.

*Acknowledgements.* A.P. thanks FAPESP (grant 02/12880−0) and CNPq (DTI grant 382.585/07−03 associated with the PCI/MCT/ON program). A.M.M. acknowledges support from FAPESP (grant 01/12589−1) and CNPq. M.B.F. acknowledges financial support from the Programme National de Physique Stellaire (PNPS-France) and the Centre National de la Recherche Scientifique (CNRS-France) for the post-doctoral grant.